\documentclass[review]{elsarticle}

\usepackage{hyperref}

\journal{Elsevier}

\begin{document}

\begin{frontmatter}

\title{Measurement of ion backflow fraction in GEM detectors}


\author[a,b]{A. Tripathy*}
\cortext[2]{Corresponding author}
\ead{alekhika.t@gmail.com}
\author[a,c]{P. K Sahu}
\author[a]{S. Swain}
\author[a]{S. Sahu}

\address[a]{Institute of Physics, HBNI, Sachivalaya Marg, P.O: Sainik School, \\Bhubaneswar-751005, Odisha, India}

\address[b]{Utkal University, Vani Vihar, Bhubaneswar-751004, Odisha, India}

\address[c]{CETMS, Siksha `O' Anusandhan University, Bhubaneswar-751030, Odisha, India}

\begin{abstract}
	
A systematic study is performed to measure the ion backflow fraction of the  GEM detectors. The effects of different voltage configurations and Ar/CO$_2$  gas mixtures, in ratios of 70:30, 80:20 and 90:10, on positive ion fraction are investigated in detail. Moreover, a comparative study is performed between single and quadruple GEM detectors.The ion current with detector effective gain is measured with various field configurations and  with three proportions of gas mixtures. The ion backflow fraction for the GEM is substantially reduced with the lower drift field.  A minimum ion backflow fraction of 18 $\%$ is achieved in the single GEM detector with Ar/CO$_2$ 80:20 gas mixture, however, a minimum ion backflow fraction of 3.5 $\%$, 3.0$\%$, and 3.8 $\%$ are obtained for a drift field of 0.1kV/cm with Ar/CO$_2$ 70:30, 80:20 and 90:10 gas mixtures, respectively for quadrupole GEM detector.  Similar values of effective gain and ion backflow fraction have been found by calculating the current from pulse height spectrum method, obtained in the Multi Channel Analyser.
   
\end{abstract}


\end{frontmatter}


\section{Introduction}


The concept of the Gas Electron Multiplier (GEM) detector was introduced in 1997  by 
Fabio Sauli \cite{sauli, sauli2016}. Some of the major advantages of the GEM detectors are 
its excellent spatial resolution, good energy resolution, high particle rate capability and long term 
stability \cite{chat2019, roy2019, patra2019, compass}.
In recent years, Gas Electron Multiplier (GEM) detector has been a preferred choice for many High
Energy Physics Experiments like COMPASS experiment at CERN  \cite{compass}, STAR, and PHENIX
experiment at RHIC \cite{star, phenix}. GEM has also been proposed for components of the International
Heavy Ion Collider \cite{ilc} and the Facility for Antiproton and Ion Research (FAIR) at GSI \cite{gsi, gem, longterm}.
These GEM detectors are also being included in the upgrade of the CMS and ALICE
experiments \cite{tpc, cms}. 
Ion backflow (IB) refers to the flow of the positive ions, that result from the
electron avalanches inside the gas, towards the drift region. Thus the electric field, that should remain
uniform for most of the cases, is distorted due to the presence of the positive ions in the drift region.

The definition of ion backflow fraction \cite{ball2014} is the ratio of the cathode current due to positive ions to the
anode current due to electrons. 
The positive ions flowing back into the drift region create a space
charge effect that affects the normal operation of the detector and results in performance degradation
of the detector \cite{ibf2019, ibf2014, ibf2004}. So, the reduction of ion backflow fraction is important for stable
detector operation. The Time Projection Chamber (TPC) and gas-filled photomultiplier (GPMT) are a few
examples in which the detector is sensitive to IB. For example, in TPC the particle tracks are
reconstructed by assuming that the drift field is uniform for estimating the x, y, z position of each voxel.
The specific energy loss (dE/dx) and the position resolution depend on the amount of IB. Similarly, in
case of GPMT the IB is an obstacle for its operation at higher gains. Therefore, an active gate electrode is
introduced between the successive GEM elements for blocking the avalanche-induced ions.

GEM can operate with different gas mixtures and can reach effective gain in the order of 10{$^2$} - 10{$^3$}, 
for a single layer amplification. To achieve further high gain, multiple layers of GEM can be coupled with the
 advantage of minimum discharge probability and lower spark rate \cite{rate, readout}. These result in 
 maximising the avalanche process in the last GEM foil. Most of the created ions are collected on the top 
 copper layer of the GEM foil during their drift path. Hence, fewer ions are capable of reaching the cathode 
 and this results in minimising the ion backflow fraction.

We have used both single and quadruple GEM detectors to study the ion backflow fraction
for these two different geometries and also with different Ar/CO$_2$ mixture.

 The paper is organised as follows: In section 2, the experimental configuration with
different configurations of detectors is described in detail. The results, for the optimisation of gain and
ion backflow fraction values for both single and quadruple GEM prototype, are discussed in section 3.
Here, the effects of the gas mixing ratio on detector performance are also presented. Finally, the
conclusion is given in the last section.


\section{Experimental Configuration}

 In this study Fe$^{55}$ X-ray source is used for measuring the effective gain and ion backflow. The X-ray source is placed at a fixed position on the top of the kapton window of the GEM detector to avoid any
spatial effect. 
 The two quantities effective gain and ion backflow fraction are defined as follows. The effective gain of the detector is defined as the anode current divided by the primary ionisation current in Eq.\ref{eq:1}. The ion backflow fraction is represented as the cathode current divided by the anode current in Eq.\ref{eq:2}. 

\begin{equation}\label{eq:1}
Effective Gain= \frac{Anode ~current}{ Primary~ionization~current   ~(R\times n \times e)}.
\nonumber
\end{equation}

\begin{equation}\label{eq:2}
 Ion~backflow~fraction = \frac{Cathode ~current}{ Anode ~current}.
\nonumber
\end{equation}

 In Eq.\ref{eq:1}, 'R' is the maximum count rate of the radioactive source, 'n' is the number of primary electrons and 'e' is the electronic charge. The maximum count rate R is measured with a scaler counter \cite{swain18}  for the different gas ratios with both detectors configuration. 
 



 The specification of  GEM foil with the standard pitch equal to 140 $\mu$m having an active area of 10 $\times$10 cm$^2$. The GEM foils used in the set-up are obtained from CERN and assembled in the cleanroom locally. The drift plane is a 50 $\mu$m kapton foil with a copper layer of 5 $\mu$m on one side which faces the top of GEM, and the readout plane is a two dimensional PCB with 128 readout pads each of area $\sim$ 9$\times$9 mm$^2$. All the pads are connected to 128 pin connectors. A sum-up board from CERN with a female lemo output is used to add the signals from all pads.

The measurements are done with the different gas ratios at Ar/CO$_2$ 70:30, 80:20, and 90:10 and constant gas flow rate, temperature, pressure, and relative humidity throughout the experiment.  

 \subsection{Experimental Set-up}

In this paper, we plan to measure ion backflow fraction for two set-ups, a single GEM and a quadruple GEM. In the beginning, the scanning is done with different voltage configurations, which is then extended by taking various gas mixture proportions. The detector architecture along with electronics arrangement for the ion backflow fraction measurements are discussed below.

The whole set-up along with the readout is placed on a copper plate and properly grounded. The schematic diagram for the GEM set-up is given in Figs. \ref{fig:1} and \ref{fig:2}.
Each electrode,  the drift plane, GEM top, and bottom are separately connected to individual common ground HV (-HV1, -HV2, -HV3 and -HV4) power supply module (CAEN N1470), the electronics set-up are as shown in Figs. \ref{fig:1} and \ref{fig:2}.  
The currents are measured from anode plane using Keithley KE6485 \cite{ke6485} pico ammeter and from cathode using PicoLogic ammeter \cite{utro2015}, which is a floating device to ensure high voltage insulation.
PicoLogic ammeter has a measuring range $\pm$125 nA, ambient temperature $25^{0}$ C,   gain 0.0039 nA/ADC count, and sampling frequency 1 kHz and is suitable for operation at a high voltage (up to $\pm$ 5kV) with respect to the ground. The full description is given in ref. \cite{utro2015} and the connection are shown in Figs. \ref{fig:1} and \ref{fig:2}. 
This arrangement gives more freedom and flexibility for operation with desired voltages. High value (10 M$\Omega$) resistors in series are used with the power supply scheme to protect GEM foil from any damage during sudden discharge. Specific voltages are set for different voltage configurations, and the respective currents are recorded from the PicoLogic ammeter. The anode current measurements are taken from a Keithley pico ammeter connected in series from the readout. All currents are measured when the 
switch 1 is connected to B and switch 2  is connected to C in Figs. \ref{fig:1} and  \ref{fig:2}.

\subsection{ Single GEM detector}

\begin{figure}[htbp]
	\centering 
	\includegraphics[width=.75\textwidth]{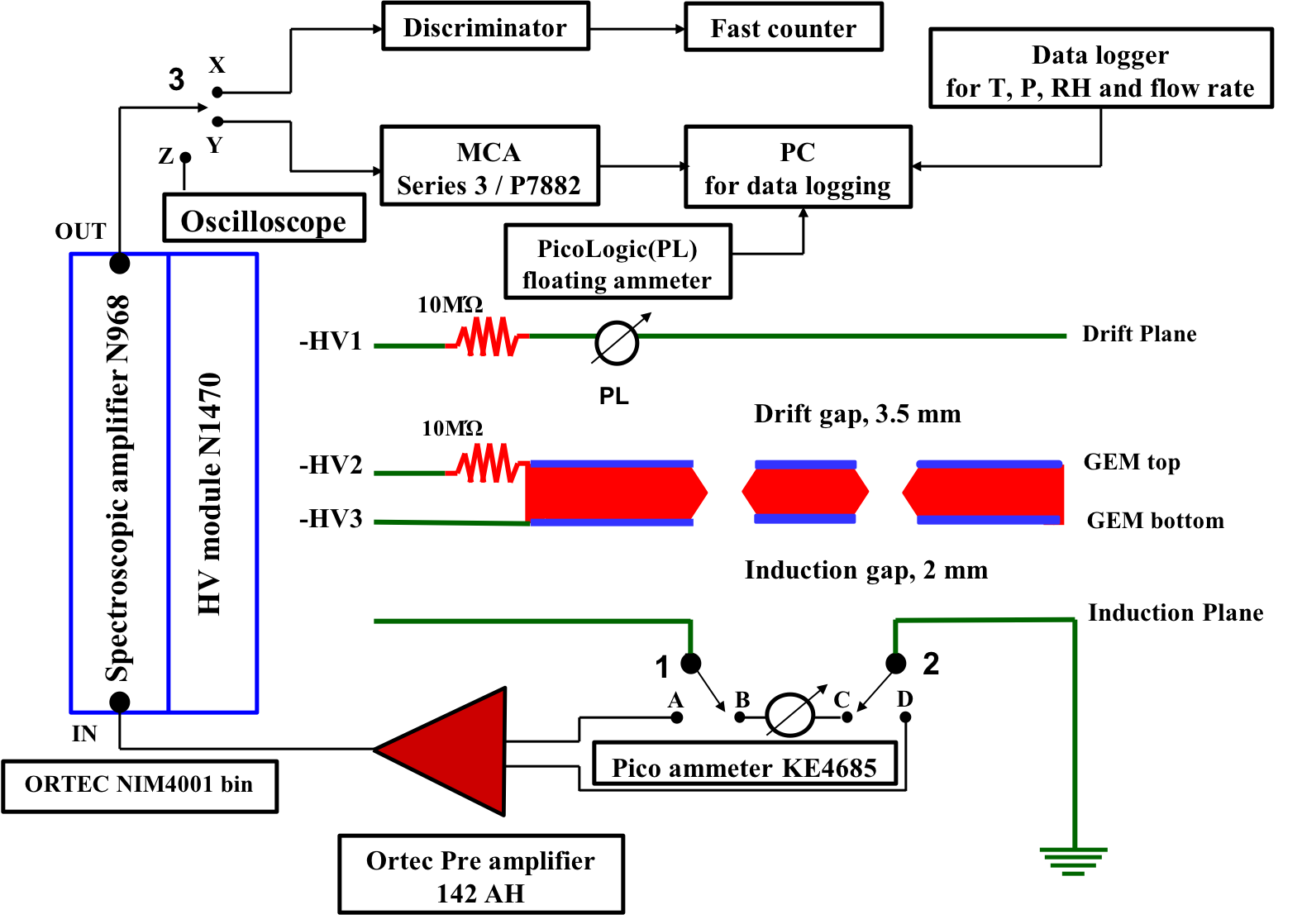}
	\caption{\label{fig:1}Schematics diagram of the single GEM detector. Voltage scheme is designed with high voltage connections at the drift plane, top and bottom layer of GEM with induction plane is being grounded. Anode current reading is taken from a Keithley Pico ammeter.}
\end{figure}

The single GEM set-up consists of a standard double mask GEM foil placed in between a drift and an
induction plane, the schematic diagram is shown in Fig. \ref{fig:1}. The GEM foil has
standard dimensions of 10cm $\times$ 10cm with holes of typically 70 $\mu$m in diameter at 140 $\mu$m pitch and has conductor on both
sides. The gap between the top of GEM foil and the drift cathode (drift plane) is 3.5mm and that between the readout anode (induction plane)
and the bottom of GEM foil is 2mm. The drift plane is a copper-clad Kapton foil
while the induction plane is realised on a printed circuit board. The whole assembly is enclosed within a
G10 frame equipped with gas inlet/outlet chamber and provided with a Kapton window. 
As our primary goal is to minimise the ion backflow fraction value for the GEM detector, we have done a scanning over different fields 
 and GEM foil amplification bias, $\Delta$V  (i.e., the voltage across top and bottom of the GEM foils).

\subsection{ Quadrupole GEM detector}

\begin{figure}[htbp]
	\centering
	\includegraphics[width=.75\textwidth]{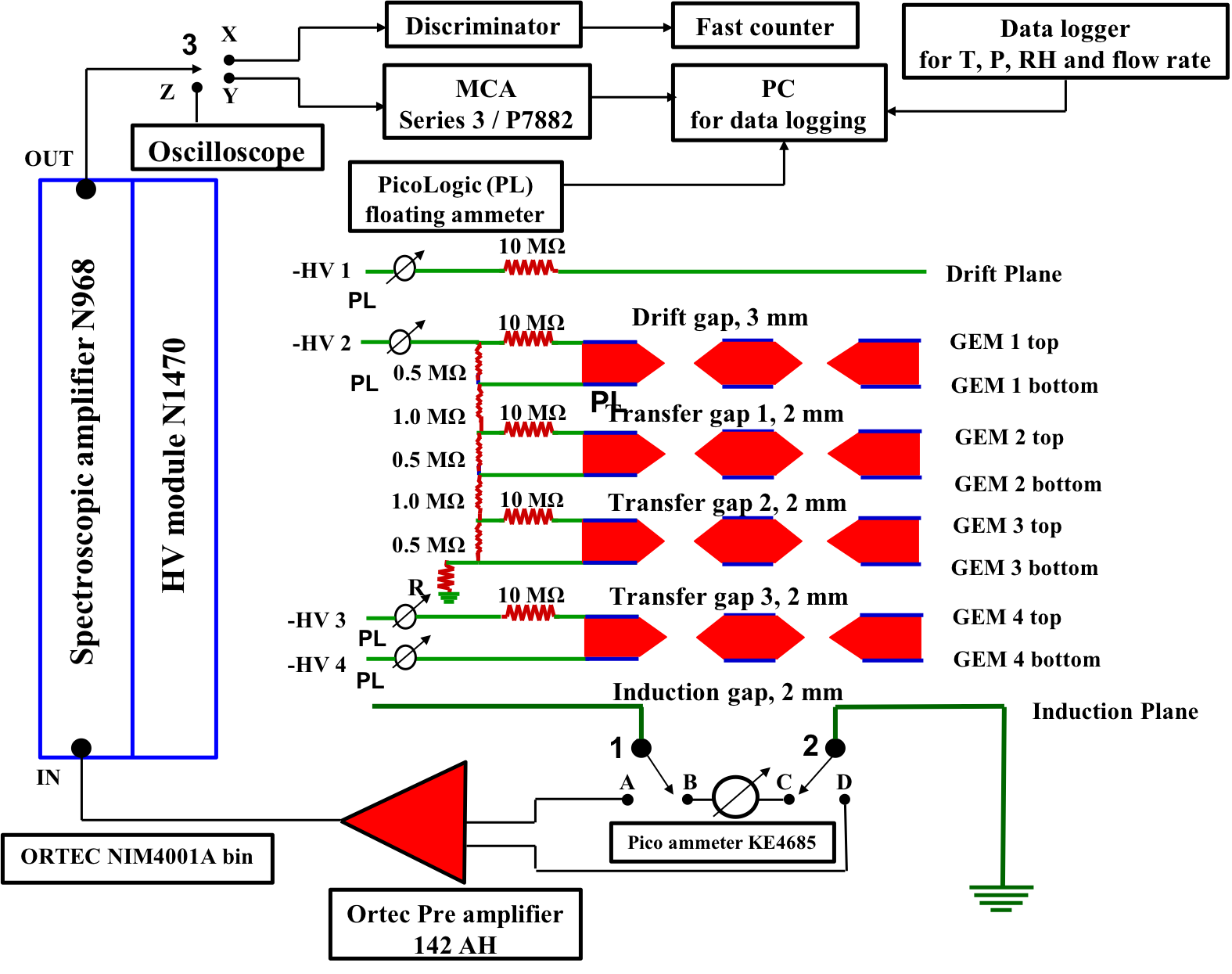}
	\caption{\label{fig:2}Schematics diagram of the quadruple GEM detector. The voltage scheme is designed with four main high voltage connections at four electrodes i.e., at the drift, top layer of GEM1, top, and bottom layers of GEM4. GEM3 bottom plane is grounded through a resistor and the induction plane is grounded and acts as the anode. Anode current reading is taken from a Keithley Pico ammeter.
}
\end{figure}

In quadrupole GEM detector four GEM foils are placed over one another with the drift plane on
the top and the readout PCB at the bottom. The readout plane consists of 128 pads with equal area over
the PCB for picking up the signals. Here also the readout plane, GEM foils and the drift plane are covered
by a G10 frame provided by a window on the top which is covered by a Kapton foil for maintaining the
gas tightness in the chamber.
The drift gap (between the drift plane and GEM1), three transfer gaps (between the GEMs), and the induction gap (between GEM4 and the induction plane) are made 3, 2, 2, 2, and 2 mm, respectively. The schematic diagram of the set-up is given in Fig. \ref{fig:2}. 

For the quad GEM set-up, the drift plane and electrodes of the GEM4 are biased with individual common ground power supply (HV N1470) -HV1, -HV3, and -HV4 as shown in Fig. \ref{fig:2}. For the distribution of correct voltages in the rest of the electrodes, a voltage divider circuit is used (-HV2). The resistance values are already mentioned in the schematic diagram. The different voltage configurations to the quad GEM are applied by pre-calculating the values of the different resistors in the detector,  thus all the fields are determined in the detector accordingly.

\subsection{ Working Principle}

 In the beginning, ultra-pure N$_2$ (99.999 $\%$) gas is passed through the detector, and gradually detector biasing is done to achieve conditioning at maximum stable operating voltage. Then Ar/CO$_2$ gas mixture is passed through the chamber with an optimised rate of flow value throughout the experiment. For both the set-ups, measurements are also done with the different gas ratios at Ar/CO$_2$ 70:30, 80:20, and 90:10 to determine the influence of the gas proportions on the ion backflow fraction values. The experiment is sensitive to ambient parameters such as gas flow rate, temperature, pressure, and relative humidity. Therefore, the corresponding gas flow rates (11 SCCM), temperature ($25^0$C), pressure (1000$\pm$3 mbar), and relative humidity ($33\%$) are recorded with a data logger\cite{datalog, sahu2018} and are maintained to be constant throughout the experiment.

As mentioned above, a sum-up board with a female lemo output connector is used for extracting the signal, measuring total counts as well as anode current from the readout plane. The electronic signal is amplified through a charge sensitive pre-amplifier and displayed in a digital oscilloscope as shown in Fig. \ref{fig:2}, the same way as shown in Fig. \ref{fig:1} for single GEM detector, when the switch 1 is connected to A, switch 2  is connected to D and switch 3 is connected to Z.  
A discriminator is used to reduce the electronic noise and it will form the detector signal to a square pulse.
The threshold voltage for single and quadrupole GEM detectors are 1.1 V and 1.28 V, respectively. 
 %
%
The extracted signal is then fed to a scaler counter to measure count rates as shown in Figs. \ref{fig:1} for single GEM detector and  Fig. \ref{fig:2} for quadrupole GEM detector when switch 1 is connected to A, switch 2  is connected to D, and switch 3 is connected to X. For the measurement of anode current, we have connected the quadrupole GEM detector output like a single GEM detector in Fig. \ref{fig:1} directly to a pico-ammeter via a lemo cable, when switches 1 and 2 are connected to B and C, respectively, in Fig. \ref{fig:2}. 
The energy spectrum of single GEM and quadrupole GEM detectors are obtained by connecting the switches 1, 2, and 3  to switches A, D, and Y, respectively, in Figs. \ref{fig:1} and \ref{fig:2}.
 Here the detector signal from the spectroscopic amplifier is fed to Multi Channel Analyser (MCA) directly with a threshold values approximately
50mV for Single GEM and 250mV for quadruple GEM detectors, respectively.

 The count rate of the GEM detector is measured as a function of $\Delta$V, which is defined as
the GEM foil amplification bias or voltage across top and bottom GEM foils.
 The GEM prototype reached a plateau region after $\Delta$V $\sim 380 $V = for all three mixed  Ar/CO$_2$ gases in proportions of 70:30, 80:20 and 90:10. The value of R is $\sim$ 48.5 kHz. The primary number of electrons is calculated and it is about 212, 217, and 222 for Ar/CO$_2$ 70:30, 80:20, and 90:10 gas ratios,  respectively. The cathode and anode currents are measured with and without the source for each voltage setting and with different gas ratios. 
The anode currents range from 0.02 nA to 0.4 nA without source and 0.08 nA to 3.6 nA with source and the cathode currents vary from 0.01 nA to 0.08 nA without source and from 0.02 nA to 0.75 nA with the source for single GEM detector.  For quadrupole GEM detector, the anode currents and cathode currents range from 0.3 nA to 6 nA and 0.08 nA to 2 nA, respectively,  without source and  from 4 nA to 90 nA and 0.6 nA to 18 nA, respectively, with the source. 
  
The drift field is varied from 0.1 kV/cm to 0.8 kV/cm. As induction field is 
important for gain, a similar scan is done for 1 to 6 kV/cm.

%
The effective gain values of the quadrupole GEM  by measuring the anode current from pico ammeter are similar to that of the  values calculated the integrated charge of the pulse height spectrum obtained by the MCA.

\begin{figure}[htbp]
	\centering 
	\includegraphics[width=.75\textwidth]{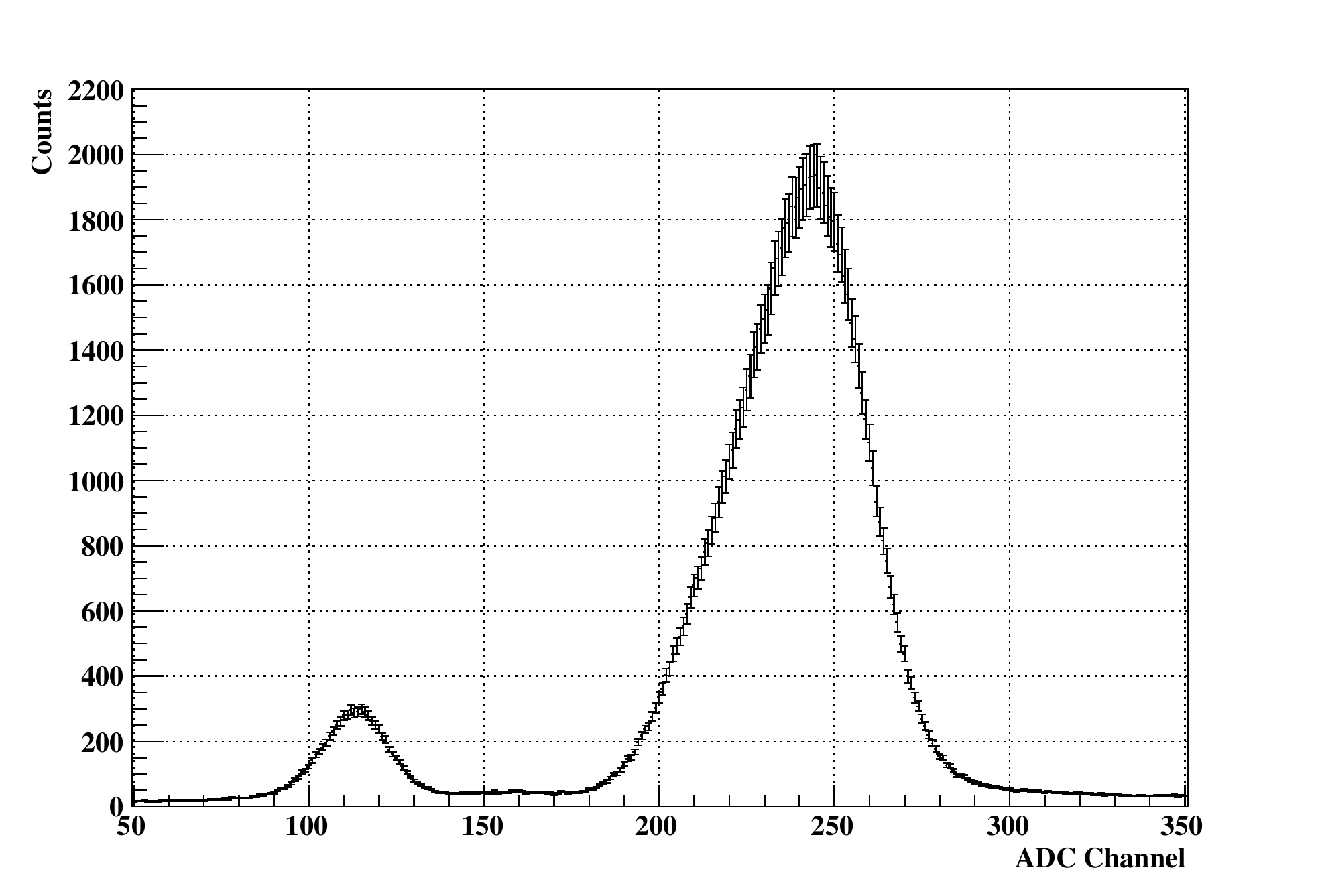}
	\caption{\label{fig:1s}Pulse height spectrum of single GEM detector for  Ar/CO$_2$ 70:30 gas ratio at $\Delta$V=410 V.  The induction field and drift field are kept constant at 4 kV/cm and 0.4 kV/cm, respectively. 
There are two peaks for Fe$^{55}$  X-ray source.  The main photopeak is corresponding to the 5.9 keV X-ray of Fe$^{55}$ and the small peak is corresponding to the 2.9 keV  Ar escape peak.} 
\end{figure}

\begin{figure}[htbp]
	\centering 
	\includegraphics[width=.75\textwidth]{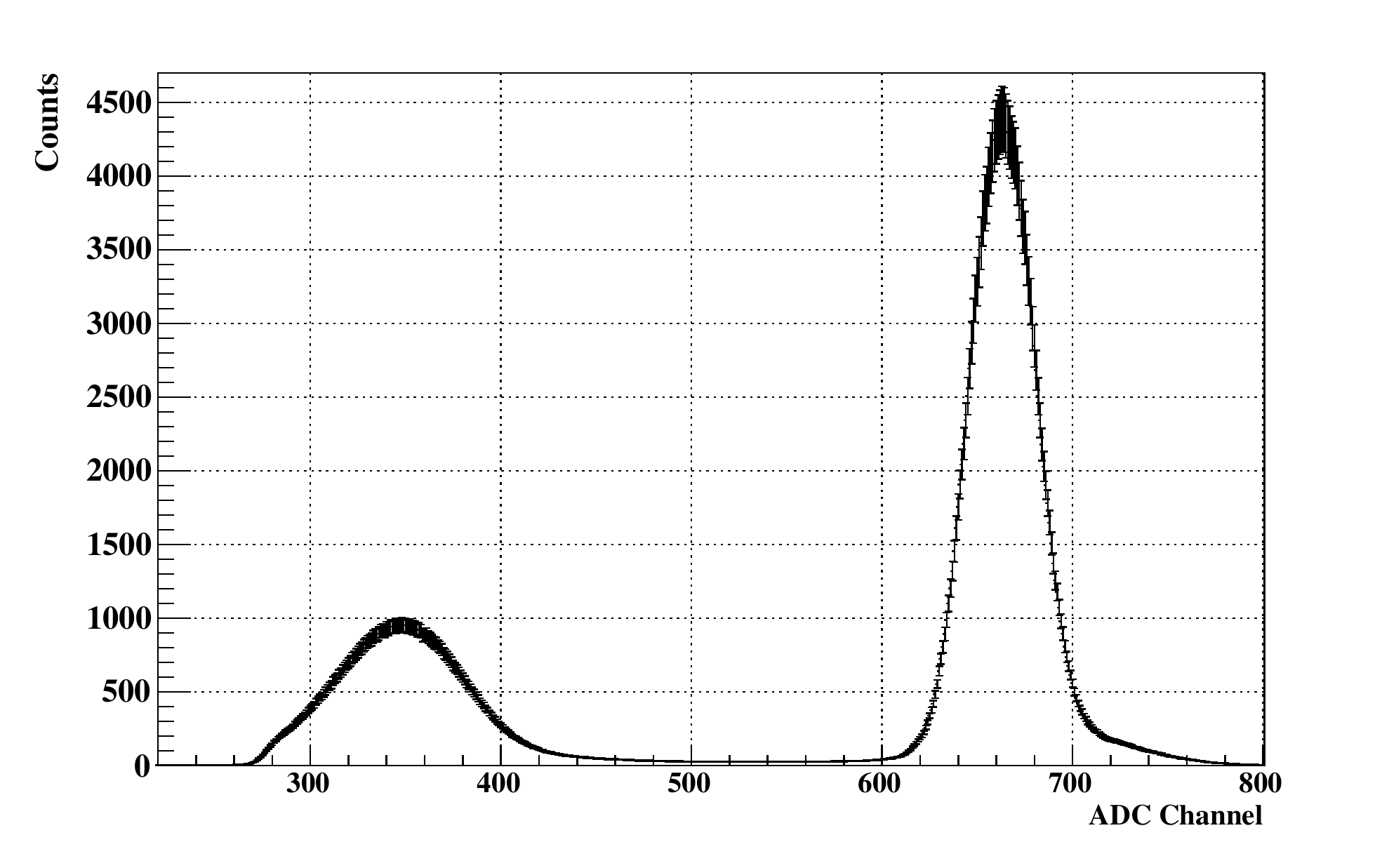}
	\caption{\label{fig:4s} Pulse height spectrum of quadrupole GEM detector for  Ar/CO$_2$ 70:30 gas mixute at $\Delta$V=320 V.  The induction field and drift field are kept constant at 4 kV/cm and 0.4 kV/cm, respectively. 
There are two peaks for Fe$^{55}$  X-ray source. 
The main photopeak is corresponding to the 5.9 keV X-ray of Fe$^{55}$ and the small peak is corresponding to the 2.9 keV  Ar escape peak.}	
\end{figure}

 We first calibrated the MCA, the calibration factor is 2.7 mV/ADC channel 
and then the signal is obtained by a sum-up board and a single input is fed to a charge sensitive preamplifier, ORTEC 142AH with 20 mV/MeV charge sensitivity. A spectroscopic amplifier CAEN N968  is connected after the preamplifier and the output signal from the amplifier is fed to the MCA to obtain the energy spectrum. 
The technical details and specification of preamplifier, ORTEC 142AH, and amplifier, CAEN N968 are given in the manuals \cite{manual1,manual2}.
The spectrum for single GEM detector with amplifier gain 50 and quadrupole GEM  detector with amplifier gain 30 are shown in Figs. \ref{fig:1s} and \ref{fig:4s}, when the switches 1, 2 and 3 are connected to switches A, D and Y, respectively in Figs. \ref{fig:1} and \ref{fig:2}.  

%
%
%


In this work, a pre-mixed Ar/CO$_2$ in 70:30 volume ratio has been used and the drift field and induction field are kept at 0.4 kV/cm and 4 kV/cm, respectively for quadrupole GEM detector. We have  determined the effective gain and ion backflow fraction in this procedure for three different values of $\Delta$V (330 V, 340 V, and 350 V) as given in  Table \ref{table:2}.

First, we calculated the total area under the pulse height spectrum and converted it into voltage using the calibration factor. This voltage is then converted into charge using the energy sensitivity factor of the pre-amplifier. This total charge per unit time gives the current and the resultant current was found to be comparable with the anode current obtained from the pico-ammeter. We estimated the effective gain from both the currents and the values agreed well given in Table \ref{table:2}. 

\section{Results}

Charge transport properties in GEM mainly depends upon detector geometry, hole configurations, gas mixtures, and different applied fields. Here we have tried to understand the detector response with a single and quadruple GEM detector. In the beginning single GEM is irradiated with 5.9 keV Fe$^{55}$ source and the measurements are taken with different applied fields and gas mixtures with the various quencher proportions.

\subsection{Single GEM detector}
 The detector is biased in such a way that the induction field is 4 kV/cm (a popular choice for ALICE TPC; low ion backflow fraction), $\Delta$V is 410 V, and drift field varies from 0.1 kV/cm to 0.6 kV/cm. 
 %
 %
%
 %
%
%
%

\begin{figure}[htbp]
	\centering 
	\includegraphics[width=.80\textwidth]{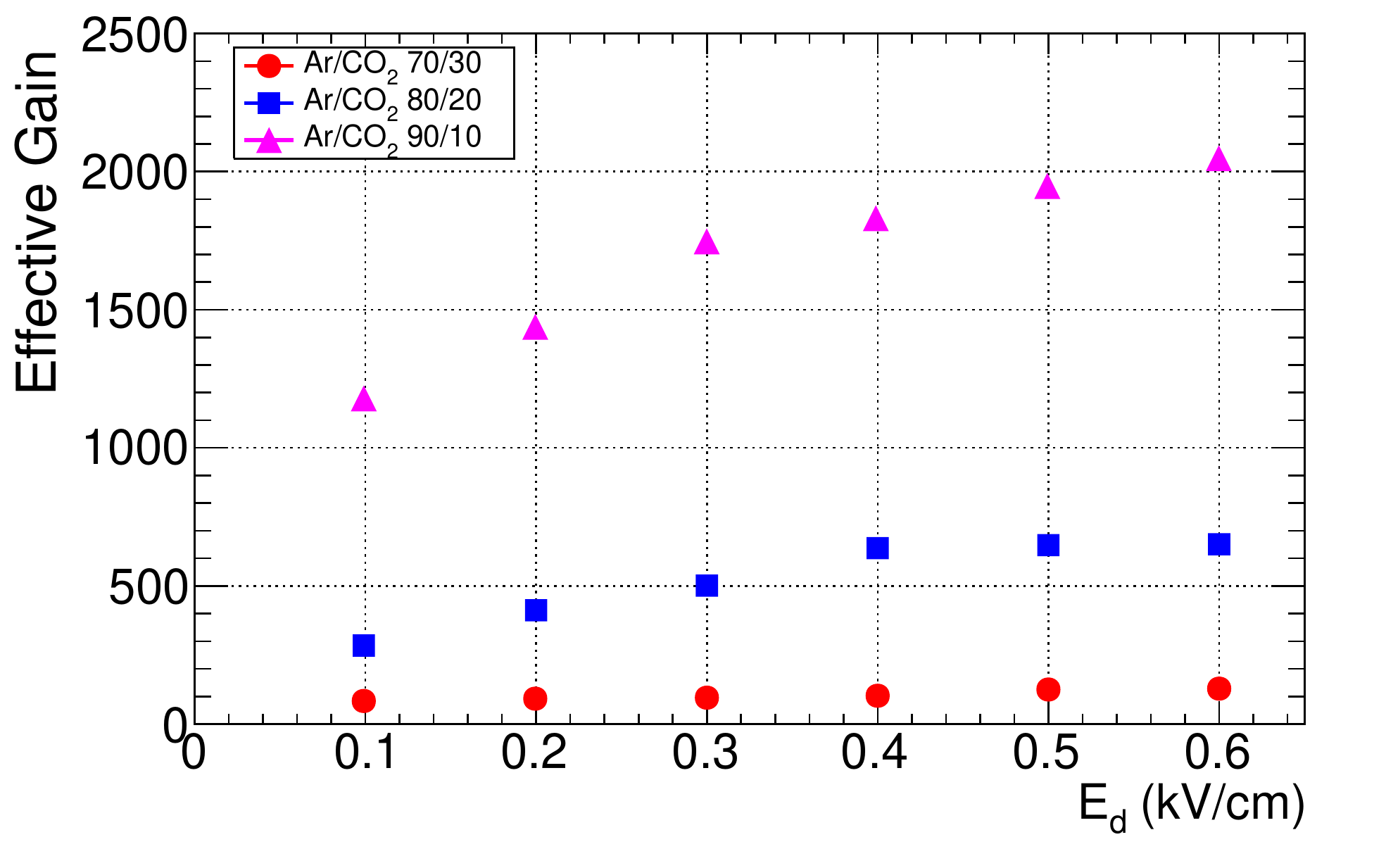}
	\caption{\label{fig:3}Effective Gain of single GEM detector as a function of the E$_d$ for different gas mixtures, E$_i$=4 kV/cm and $\Delta$V= 410 V were kept constant. }
\end{figure}

\begin{figure}[htbp]
	\centering 
	\includegraphics[width=.80\textwidth]{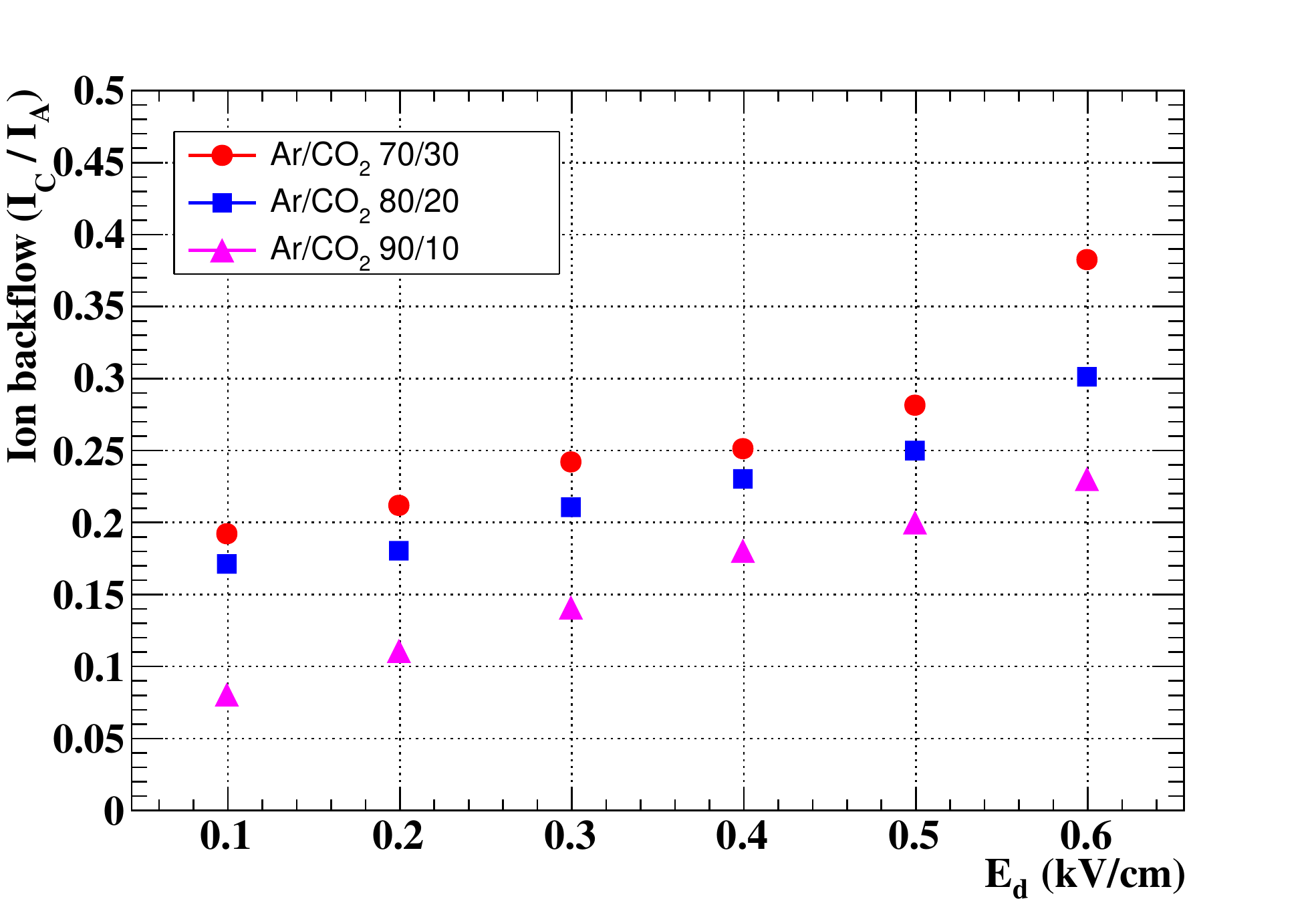}
	\caption{\label{fig:4}Ion backflow fraction of single GEM detector as a function of E$_d$ for different gas mixtures, E$_i$=4 kV/cm and $\Delta$V=410 V  were kept constant. }
\end{figure}

 %
The effective gain is calculated from the corresponding anode current, and a comparison is made for three mixed Ar/CO$_2$ 70:30, 80:20, and 90:10 gas ratios as shown in Fig. \ref{fig:3}.  For the same voltage settings, the cathode current is noted and ion backflow fraction is calculated. Fig. \ref{fig:4} shows the effects of drift field $E_d$ on ion backflow fraction for the three different gas mixtures.
Effective gain increases as a  function of induction fields. The effective gains reach to a  plateau region after $E_i > $ 4 kV/cm, 
 for all three Ar/CO$_2$  70:30, 80:20 and 90:10 gas ratios. 
%


\begin{figure}[htbp]
	\centering 
	\includegraphics[width=.80\textwidth]{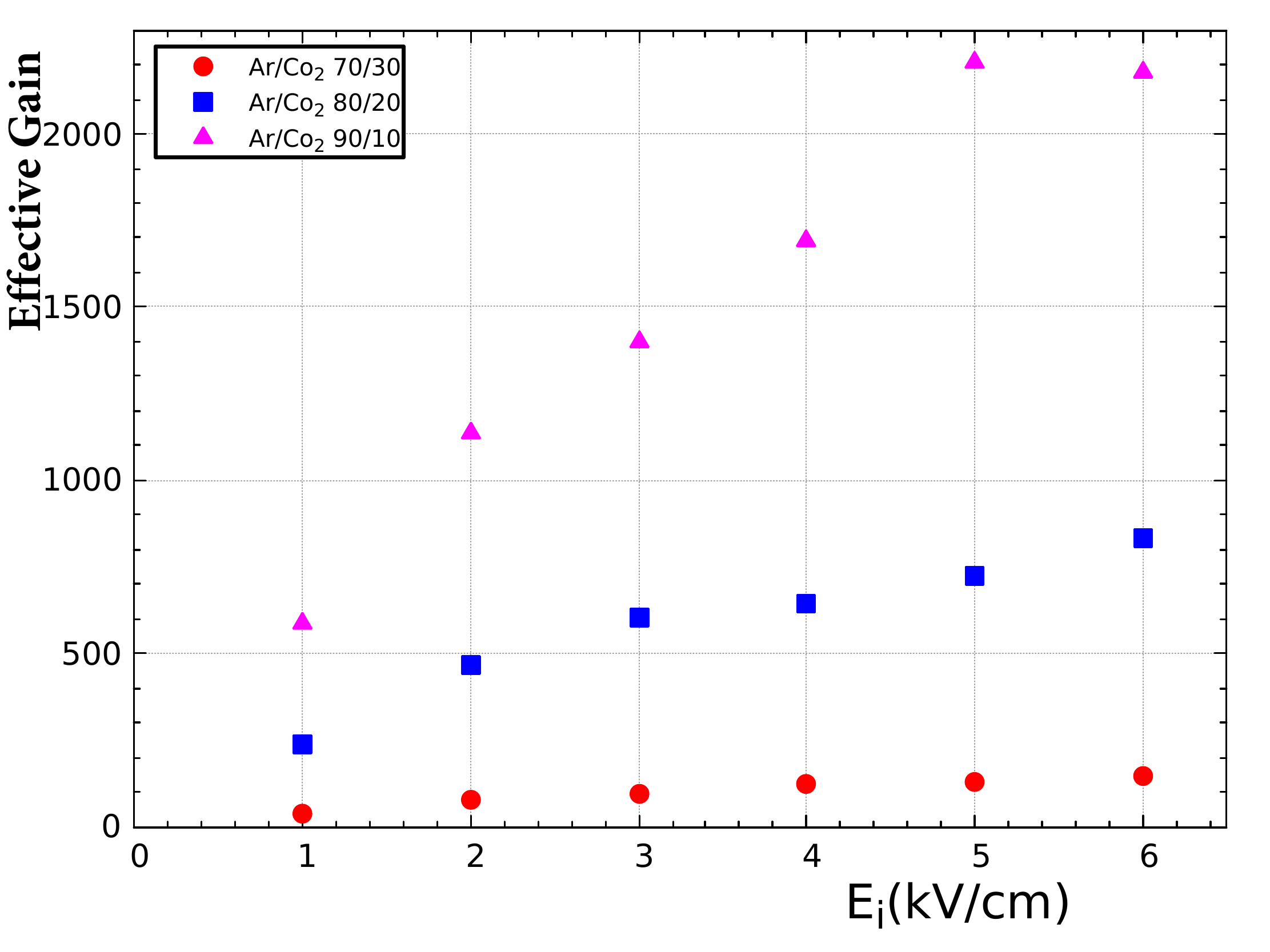}
	\caption{\label{fig:5}Effective Gain of single GEM detector as a function of the E$_i$ for three different gas mixtures, where E$_d$=0.4 kV/cm and $\Delta$V=410 V were kept constant.}
\end{figure}

\begin{figure}[htbp]
	\centering 
	\includegraphics[width=.80\textwidth]{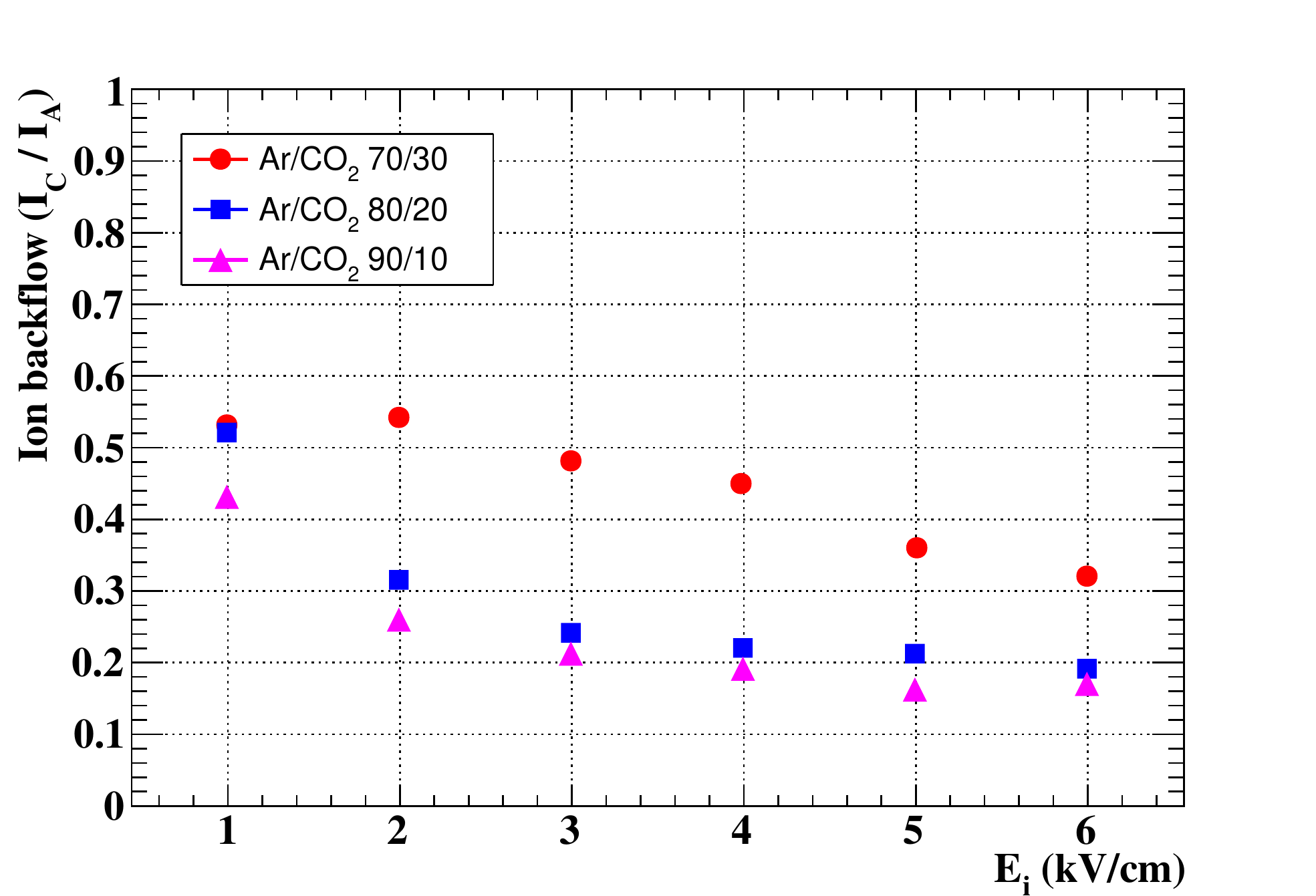}
	\caption{\label{fig:6}Ion backflow fraction of a single GEM detector as a function of E$_i$ for different gas mixtures, E$_d$=0.4 kV/cm  and $\Delta$V= 410 V were kept constant. }
\end{figure}

\begin{figure}[htbp]
	\centering 
	\includegraphics[width=.80\textwidth]{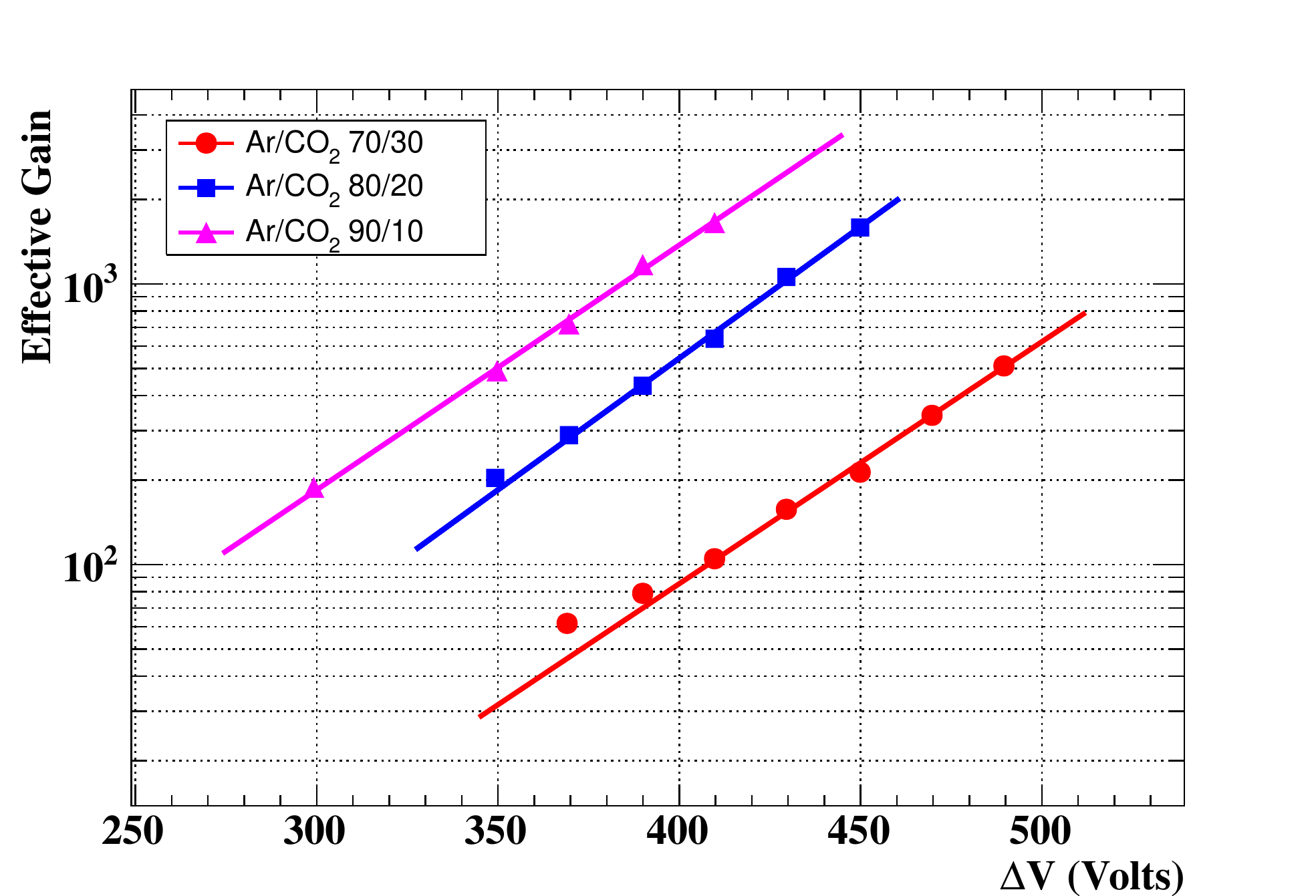}
	\caption{\label{fig:7}Effective Gain of single GEM detector as a function of $\Delta$V for constant values of $E_i$= 4 kV/cm and $E_d$=0.4 kV/cm.}
\end{figure}

\begin{figure}[htbp]
	\centering 
	\includegraphics[width=.80\textwidth]{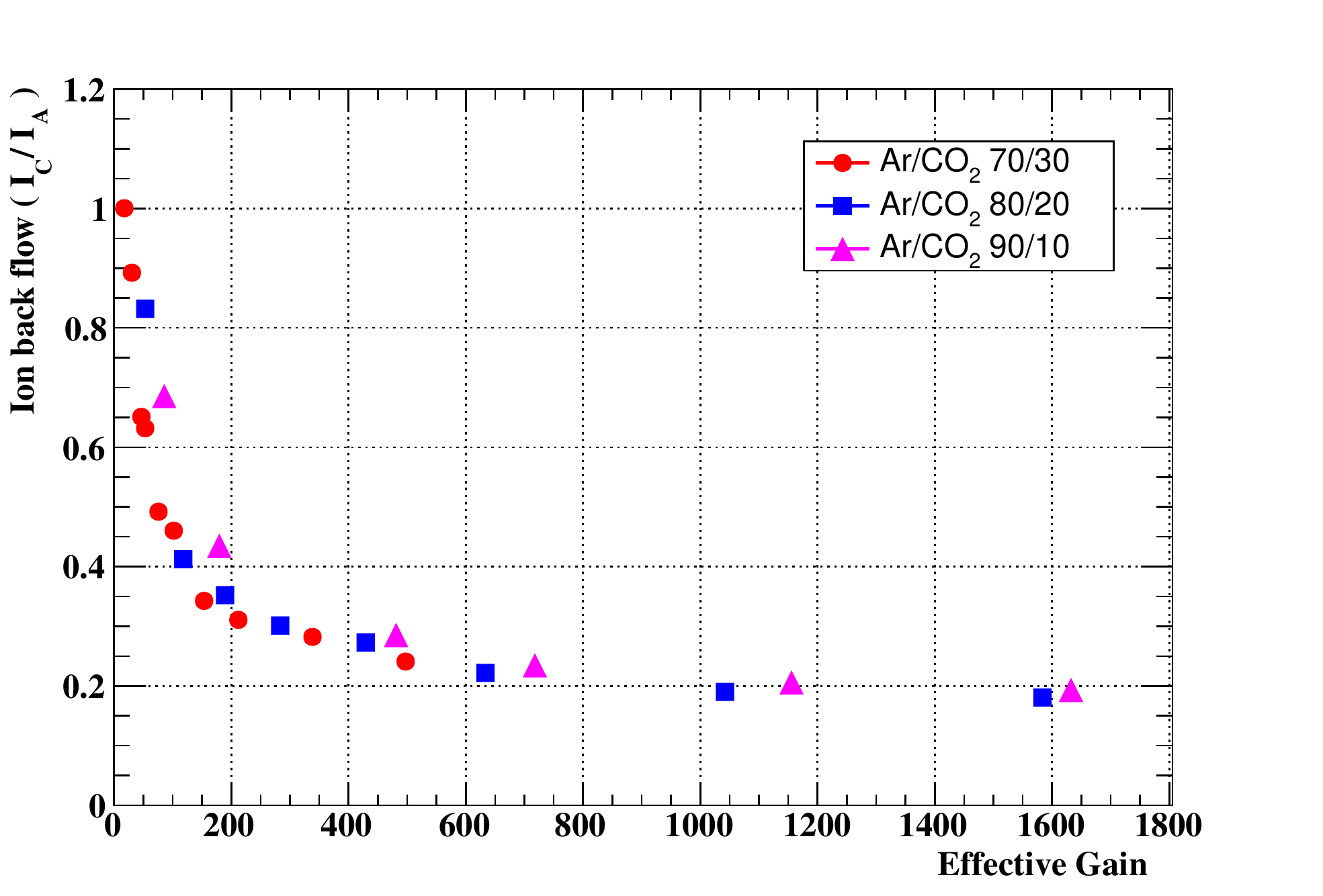}
	\caption{\label{fig:8}Ion backflow fraction of single GEM detector with effective gain keeping $\Delta$V=410 V, and $E_d$=0.4 kV/cm by varying the $E_i$ value maximum up to 6 kV/cm.}
\end{figure}
 
 Here, we see that effective gain is 
 increasing with $E_d$ for Ar/CO$_2$ 70:30, 80:20 and 90:10 gas ratios.  It increases slowly and  
 saturates after  $E_d$=0.4 kV/cm for all gas ratio and on the other hand, ion backflow fraction rises linearly with E$_d$ \cite{ibf2019}. It is because with increasing drift field more electric field lines end on the drift plane giving rise to a higher value of cathode current. This, in turn, increases the ion backflow fraction. 

An optimum value of the drift field is desired to create the primary number of electrons and push them into the GEM holes. But here it is observed that, after a particular value of the field, the effective gain is 
insensitive to change in the drift field. When we consider different Ar/CO$_2$ gas mixtures, the effective gain is maximum for a 90:10 ratio due to the maximum content of ionisation component. On the same argument, the anode current value increases with Ar content, we get minimum ion backflow fraction for this ratio. 

The effective gain and ion backflow fraction behaviour are also observed with changing the induction field by keeping the drift field and GEM voltage constant. Now the detector is biased in such a way that the drift field is 0.4 kV/cm, $\Delta$V is 410 V, and the induction field varies from 1 kV/cm to 6 kV/cm.  To make a comparison of the performances of different gas mixtures, the effective gain and ion backflow fraction are displayed as a function of induction field E$_i$ in Fig. \ref{fig:5} and Fig. \ref{fig:6}, respectively.

The effective gain of the detector is calculated and plotted with applied GEM voltage $\Delta$V for different gas mixtures by keeping E$_d$ = 0.4 kV/cm and E$_i$ = 4kV/cm constant. It is observed that effective gain increases exponentially with $\Delta$V and maximum effective gain is recorded for Ar/CO$_2$ 90:10 ratio due to less quencher content, as is shown in Fig. \ref{fig:7}. Finally, to optimise our detector set-up, we have plotted the ion backflow fraction with corresponding effective gain value by keeping $\Delta$V =410 V for three gas ratios. This can be seen in Fig. \ref{fig:8}. For Ar/CO$_2$ 70:30 ratio, the maximum effective gain value is approximately  500, as we have varied the $E_i$ value maximum up to 6 kV/cm. The decreasing trend of the ion backflow fraction as a function of effective gain is the same as given in Ref.\cite{ibf2019}.  A minimum ion backflow fraction of 18$\%$ is achieved in the single GEM set-up with Ar/CO$_2$ 80:20 gas at effective gain around 1500 for this set-up. 
%
%

\subsection{Quadruple GEM detector}
In the previous section, the operation of a single GEM prototype has been investigated, and the role of fields along with gas mixture in different ratios has been discussed. In the present section, the 
GEM prototype is operated with different voltage biasing schemes and the effective gain with ion backflow fraction values are optimised. The same 5.9 keV Fe$^{55}$ radioactive source is used. 
 The primary purpose of our study is to optimise the values of $\Delta$V, $E_i$, $E_d$ and effective gain  of the quadruple detector for the lowest possible value of the ion backflow fraction for 
 three Ar/CO$_2$ gas ratios. 

\begin{table}

	\begin{center}
		\begin{tabular}{||c c c c||} 
			\hline
			Gas mixture & & & 
			\\ Ar/CO$_2$  & E$_d$ (kV/cm) & E$_i$ (kV/cm) & $\Delta$ V (V)\\
			\hline\hline
			70:30 & 0.1 - 0.8 & 3 - 6 & 270 - 330 \\ 
			\hline
			80:20 & 0.1 - 0.8 & 3 - 6 & 250 - 310 \\
			\hline
			90:10& 0.1 - 0.8 & 3 - 6 &  220 - 280 \\
			\hline
		\end{tabular}
		\caption{\label{table:1}Field configuration for quadruple GEM detector}
		
	\end{center}
\end{table}

\begin{table}

	\begin{center}
		\begin{tabular}{||c c c||} 
			\hline
		  Effective Gain  & IBF($\%$) & $\Delta$V (V)\\
			\hline\hline
			39000 & 18 & 330 \\ 
			\hline
			42000 & 13 & 340 \\
			\hline
			42000& 9 &  350  \\
			\hline
		\end{tabular}
		\caption{\label{table:2}Effective Gain and ion backflow fraction values of quadruple GEM detector at different $\Delta$V for  
                 E$_d$=0.4 kV/cm and E$_i$=4 kV/cm by pulse height method.}
		
	\end{center}
\end{table}

By looking at the results of a single GEM, we first scan the detector with field variation which is then continued to study with different gas mixtures. In the beginning, the detector is biased as per the voltage settings, and then effective gain and ion backflow fraction value are calculated. The drift and induction fields are kept as 0.4kV/cm and 4 kV/cm, respectively. The selected field ranges for the detector with different gas mixtures are listed in Table \ref{table:1}. 

\begin{figure}[htbp]
   	\centering 
   	\includegraphics[width=.80\textwidth]{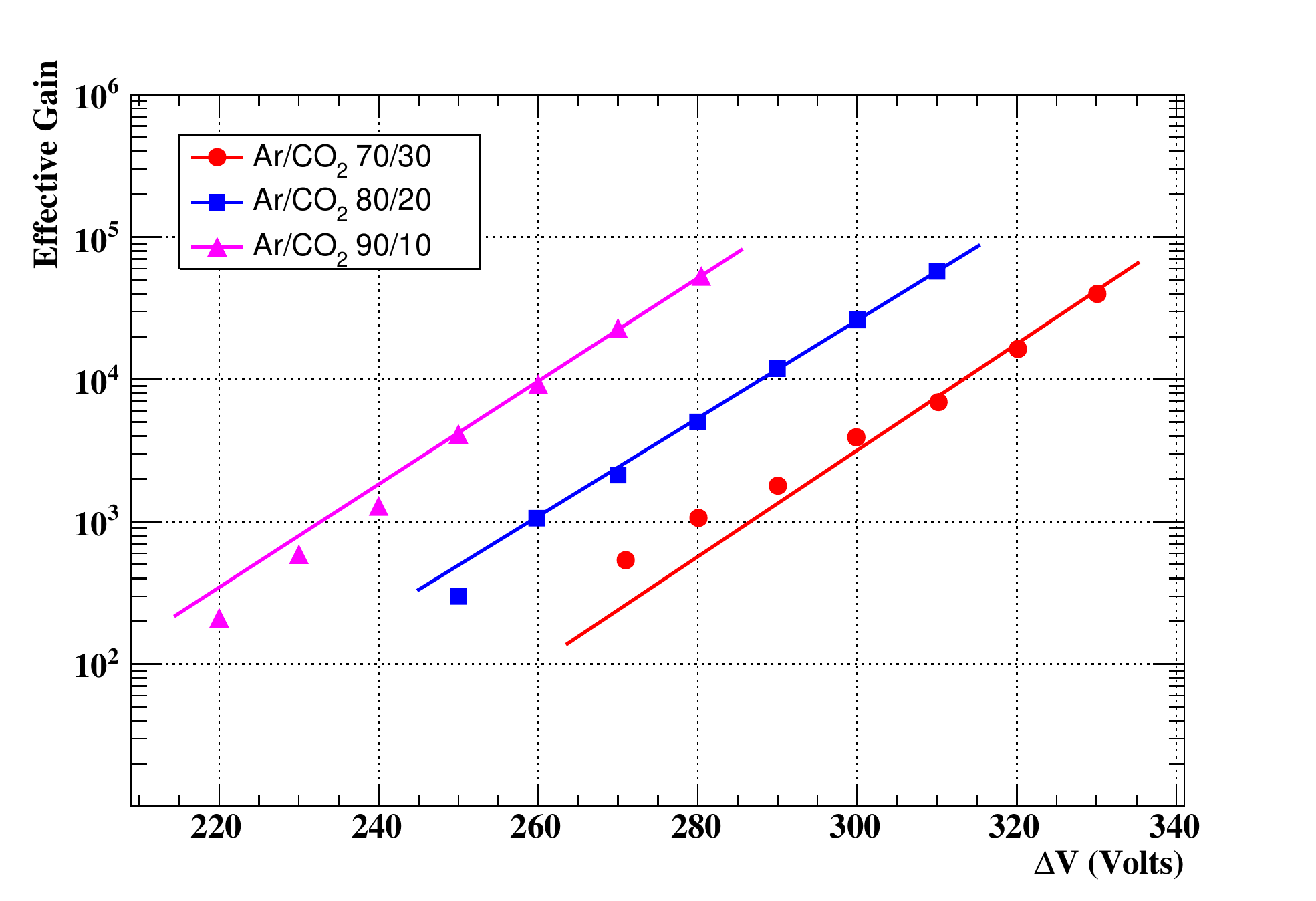}
   	\caption{\label{fig:9}Effective Gain of quadruple GEM with  $\Delta$V voltage at different mixed gas ratio, E$_d$ and E$_i$ are being constant at 0.4kV/cm and 4kV.cm, respectively.}
   \end{figure}

Also, the effective gain calculation from the pulse height spectrum obtained from the MCA is done with a 70:30 gas ratio and the drift and induction fields are kept at 0.4 kV/cm and 4 kV/cm, respectively, as shown in the Table \ref{table:2}. The effective gain increases exponentially with increasing $\Delta$V. The qualitative behavior is the same as that estimated by the current measurement method.
%
%
%
%
The GEM voltages $\Delta$V are kept 320 V, 300 V and 270 V for Ar/CO$_{2}$ gas ratio 70:30, 80:20 and 90:10, respectively.
 The effective gain increases exponentially with increasing $\Delta$V for all gases, and the plot is shown in Fig. \ref{fig:9}. 
 %
 %
 The detector is maintained at a constant effective gain, and the variation of ion backflow fraction value is then investigated. For the next measurements, the detector is operated with voltages above 10$^4$ gain. For the higher proportion of ionising gas, the detector is restricted for the working in higher $\Delta$V regions. So, for Ar/CO$_2$  in 80:20 and 90:10 ratio, higher GEM voltage $\Delta$V is not possible due to the continuous occurrence of sparks. At a constant GEM voltage $\Delta$V, the effective gain is maximum for Ar/CO$_2$ 90:10 ratio due to more ionisation and a higher proportion of Ar. 

 %

 \begin{figure}[htbp]
 	\centering 
 	\includegraphics[width=.75\textwidth]{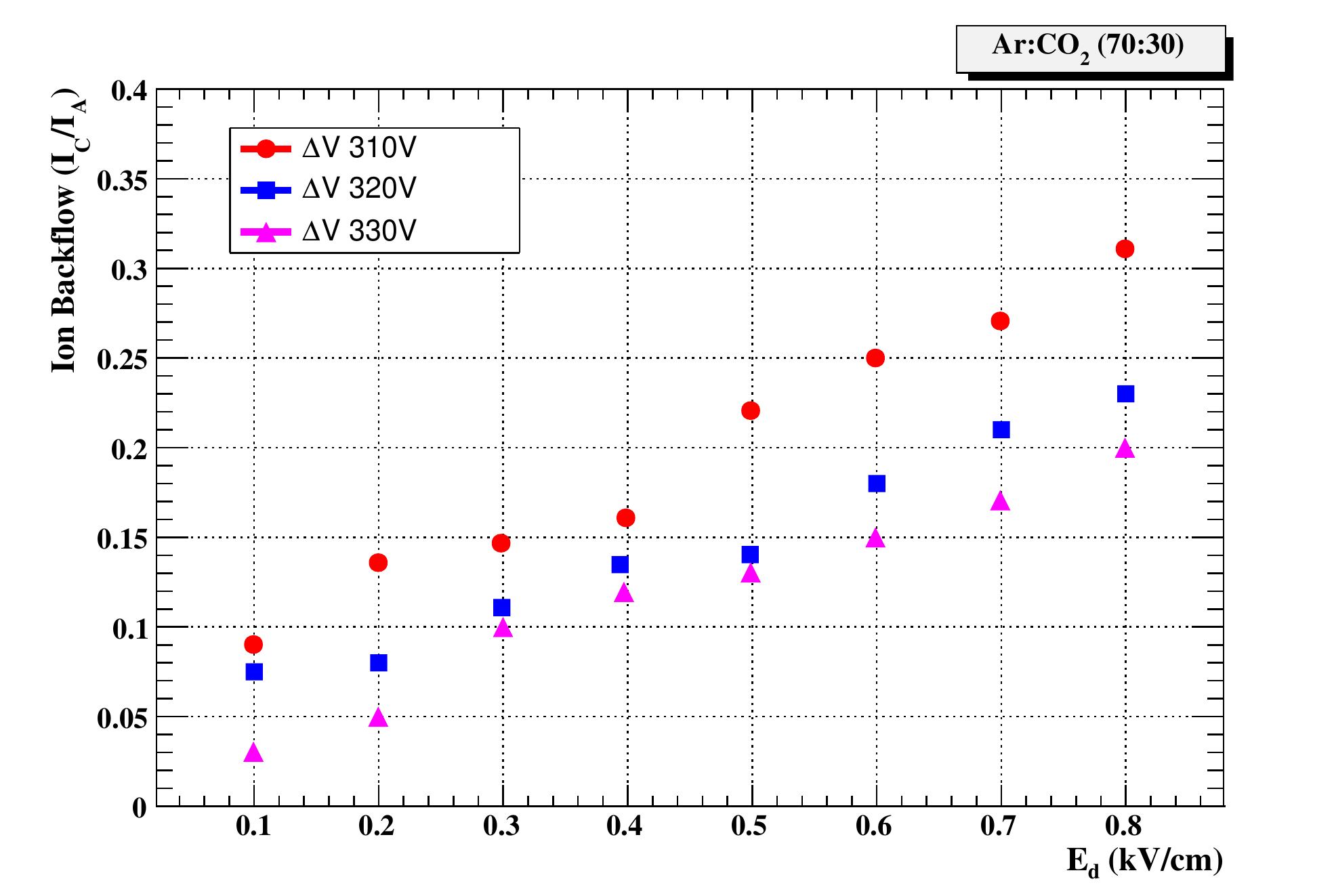}
 	\caption{\label{fig:10}Ion backflow fraction of quadrupole GEM detector with variation of drift field E$_d$ at different GEM voltages $\Delta$V for  Ar/CO$_2$ 70:30. The induction field is kept constant at 4 kV/cm.}
 \end{figure}
 
  \begin{figure}[htbp]
 	\centering 
 	\includegraphics[width=.80\textwidth]{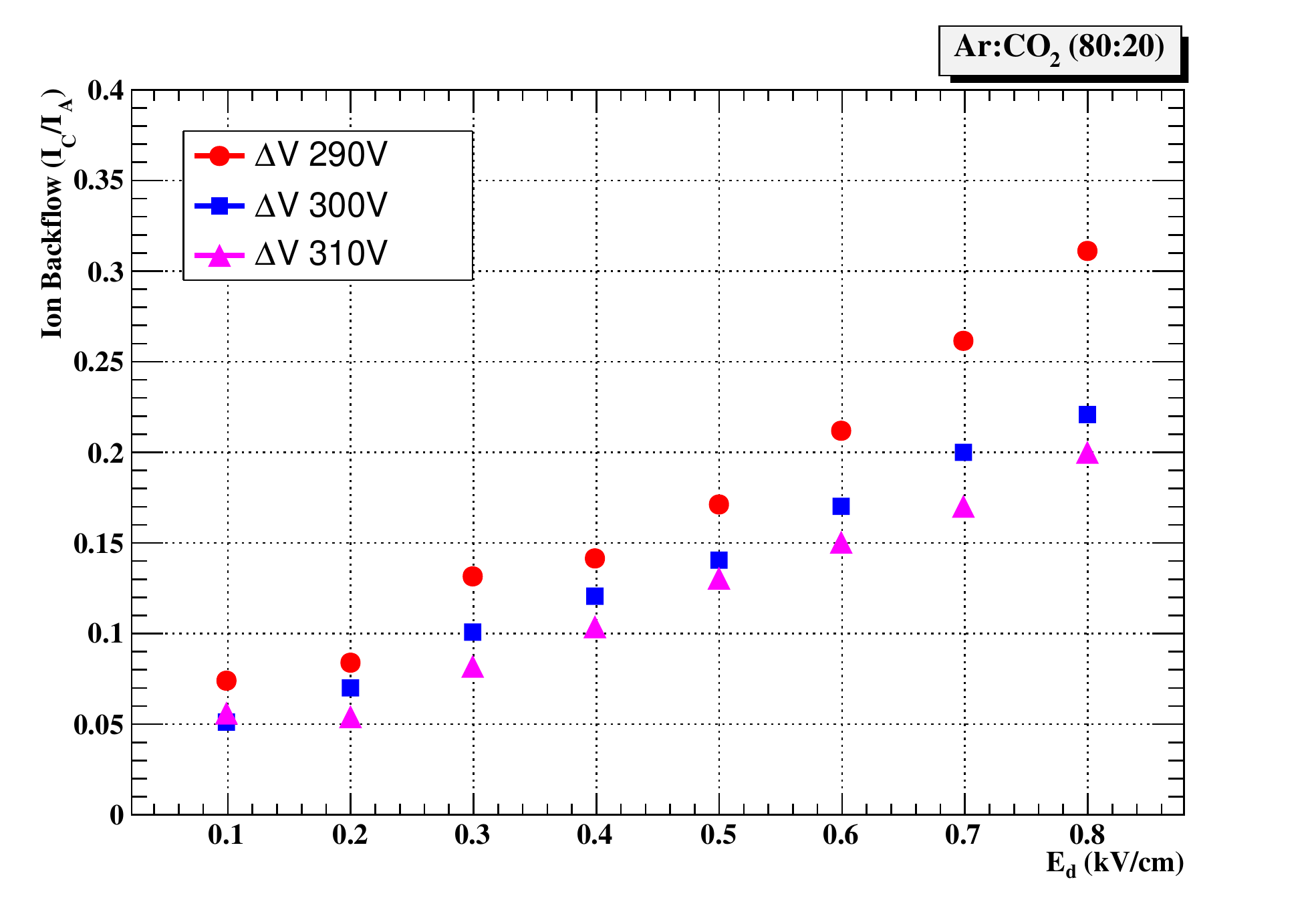}
 	\caption{\label{fig:11}Ion backflow fraction of quadrupole GEM detector with variation of drift field E$_d$ at different GEM voltages $\Delta$V for  Ar/CO$_2$ 80:20. The induction field is kept constant at 4 kV/cm.}
 \end{figure}

 \begin{figure}[htbp]
	\centering 
	\includegraphics[width=.80\textwidth]{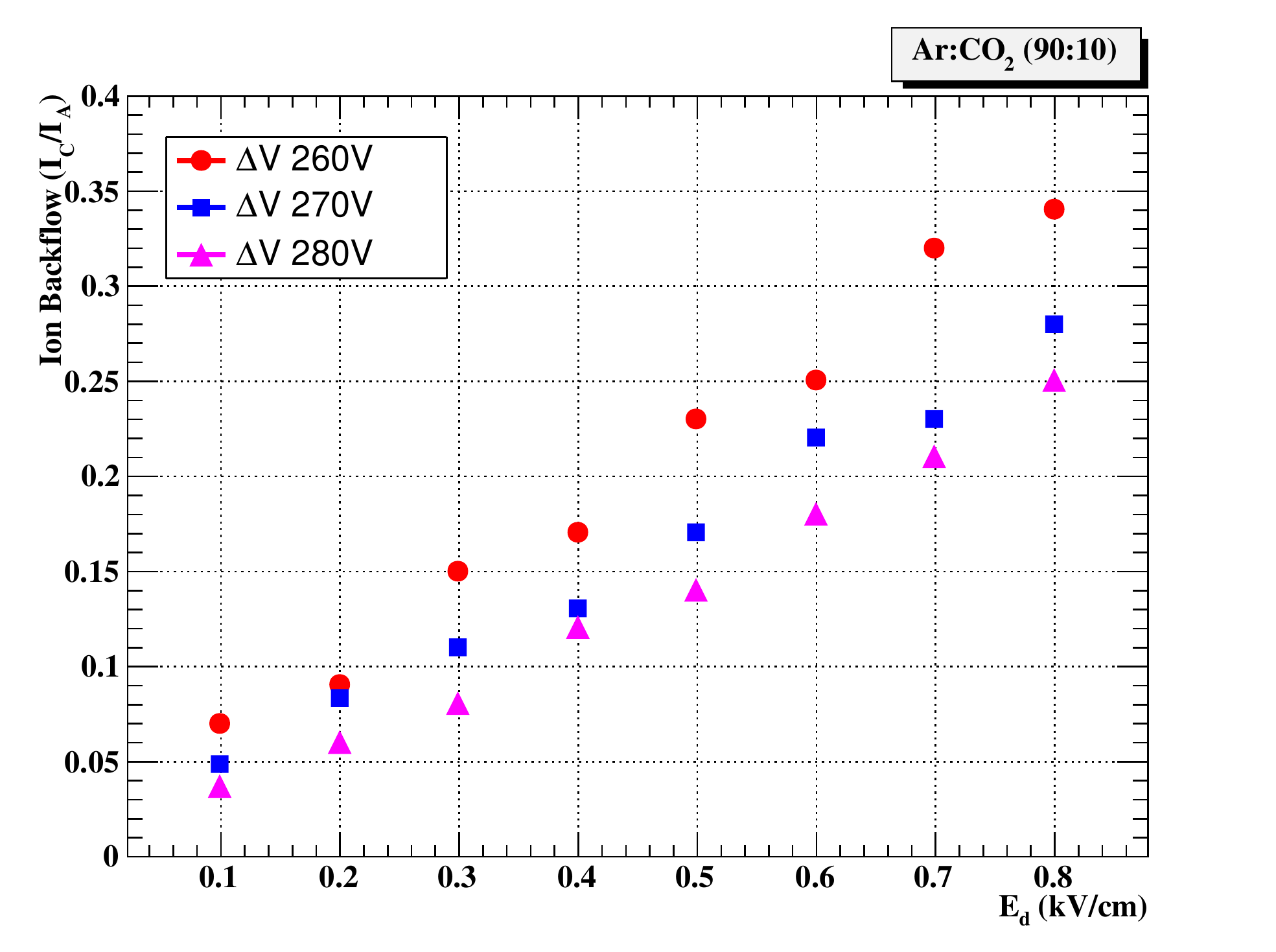}
	\caption{\label{fig:12}Ion backflow fraction of quadrupole GEM detector with variation of drift field E$_d$ at different GEM voltages $\Delta$V for  Ar/CO$_2$ 90:10. The induction field is kept constant at 4 kV/cm.}
\end{figure}

\begin{figure}[htbp]
	\centering 
	\includegraphics[width=.80\textwidth]{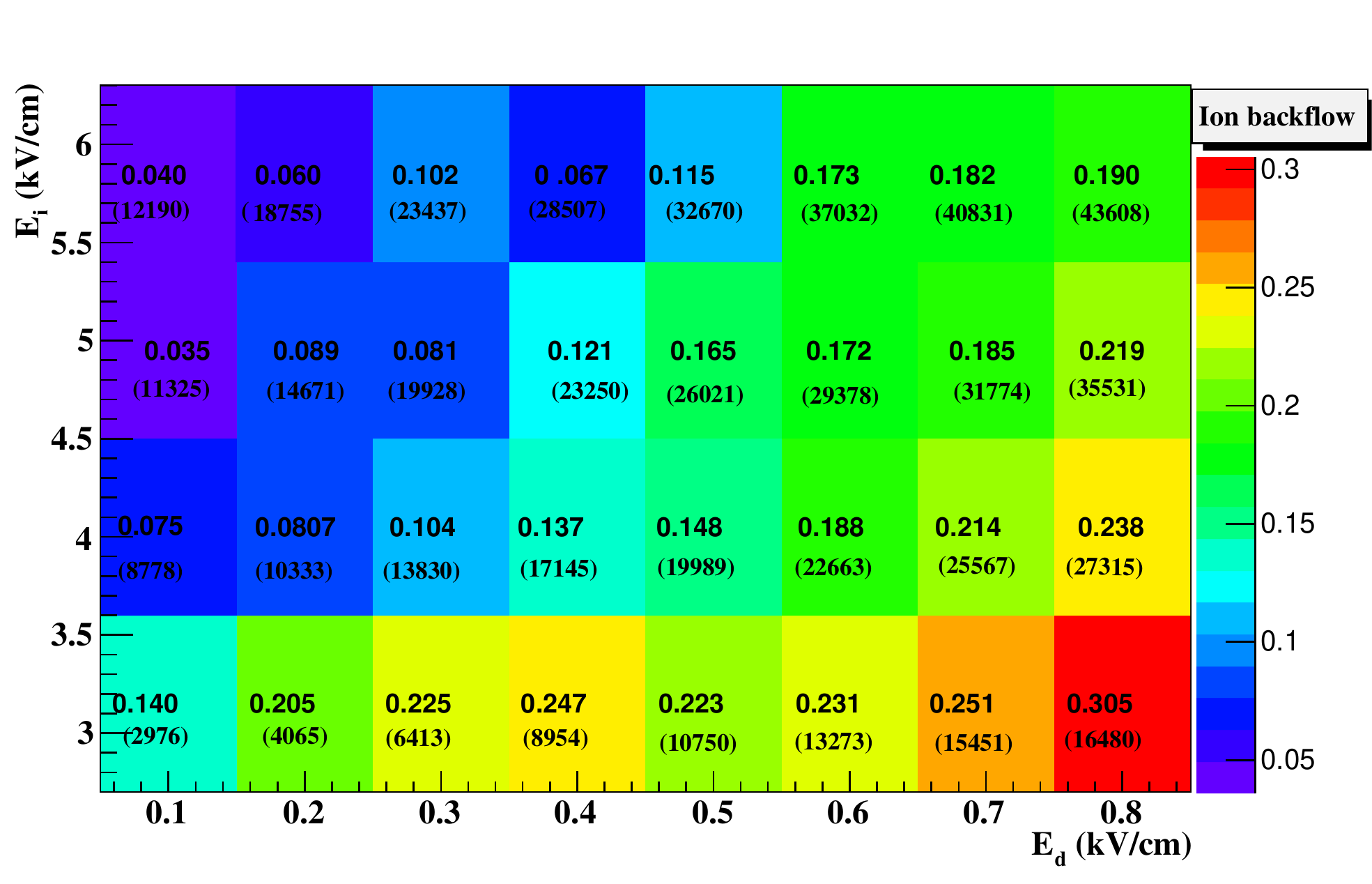}
	\caption{\label{fig:13}Ion backflow fraction of quadrupole GEM detector measured with scanning over E$_d$ and E$_i$ at constant GEM voltage $\Delta$V = 320 V for Ar/CO$_2$ 70:30. The numbers in parenthesis represent the effective gain.}
\end{figure}

\begin{figure}[htbp]
	\centering 
	\includegraphics[width=.80\textwidth]{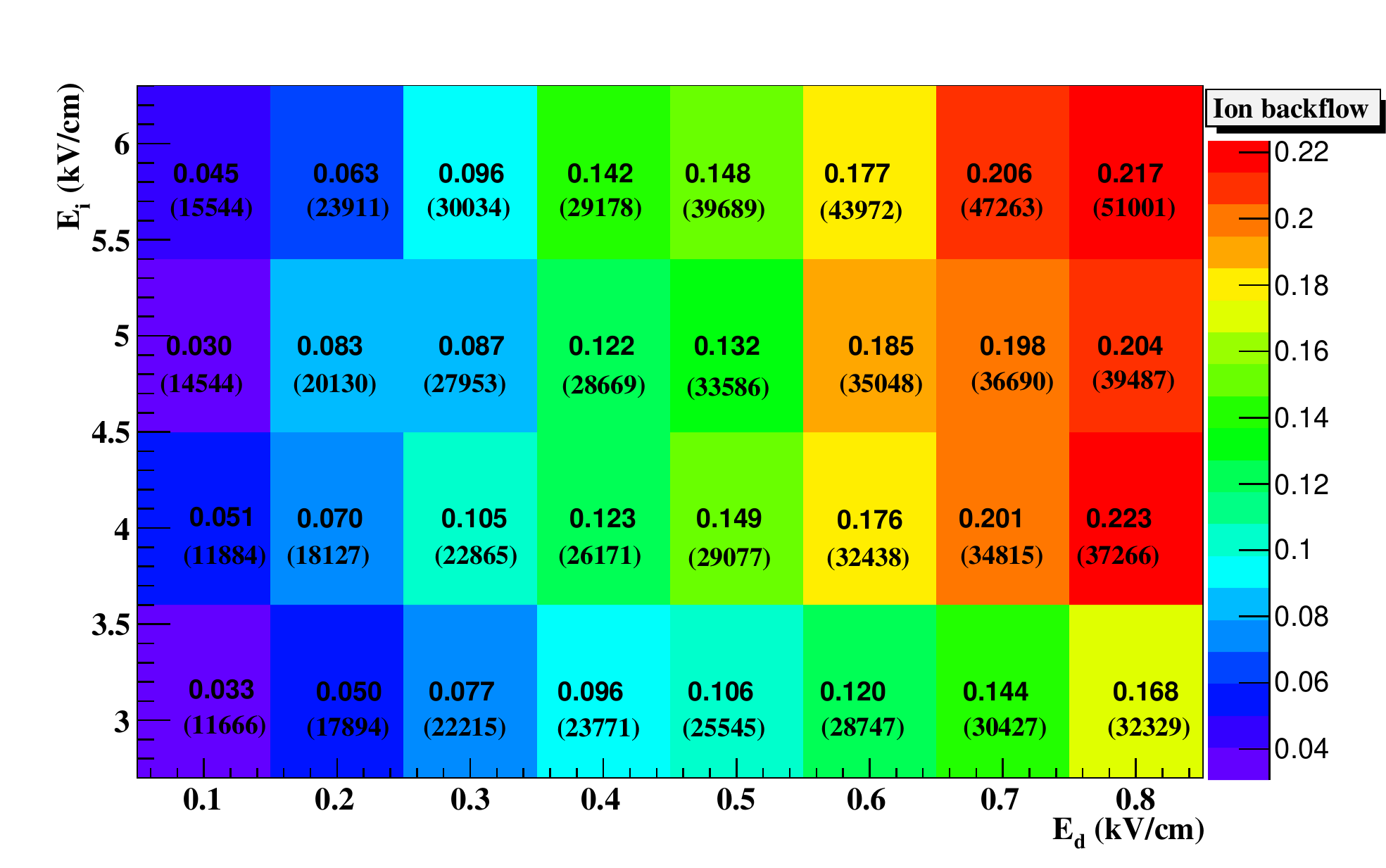}
	\caption{\label{fig:14}Ion backflow fraction of quadrupole GEM detector measured with scanning over E$_d$ and E$_i$ at constant GEM voltage $\Delta$ V = 300 V for Ar/CO$_2$ 80:20. The numbers in parenthesis represent the effective gain.}
\end{figure}

\begin{figure}[htbp]
	\centering 
	\includegraphics[width=.80\textwidth]{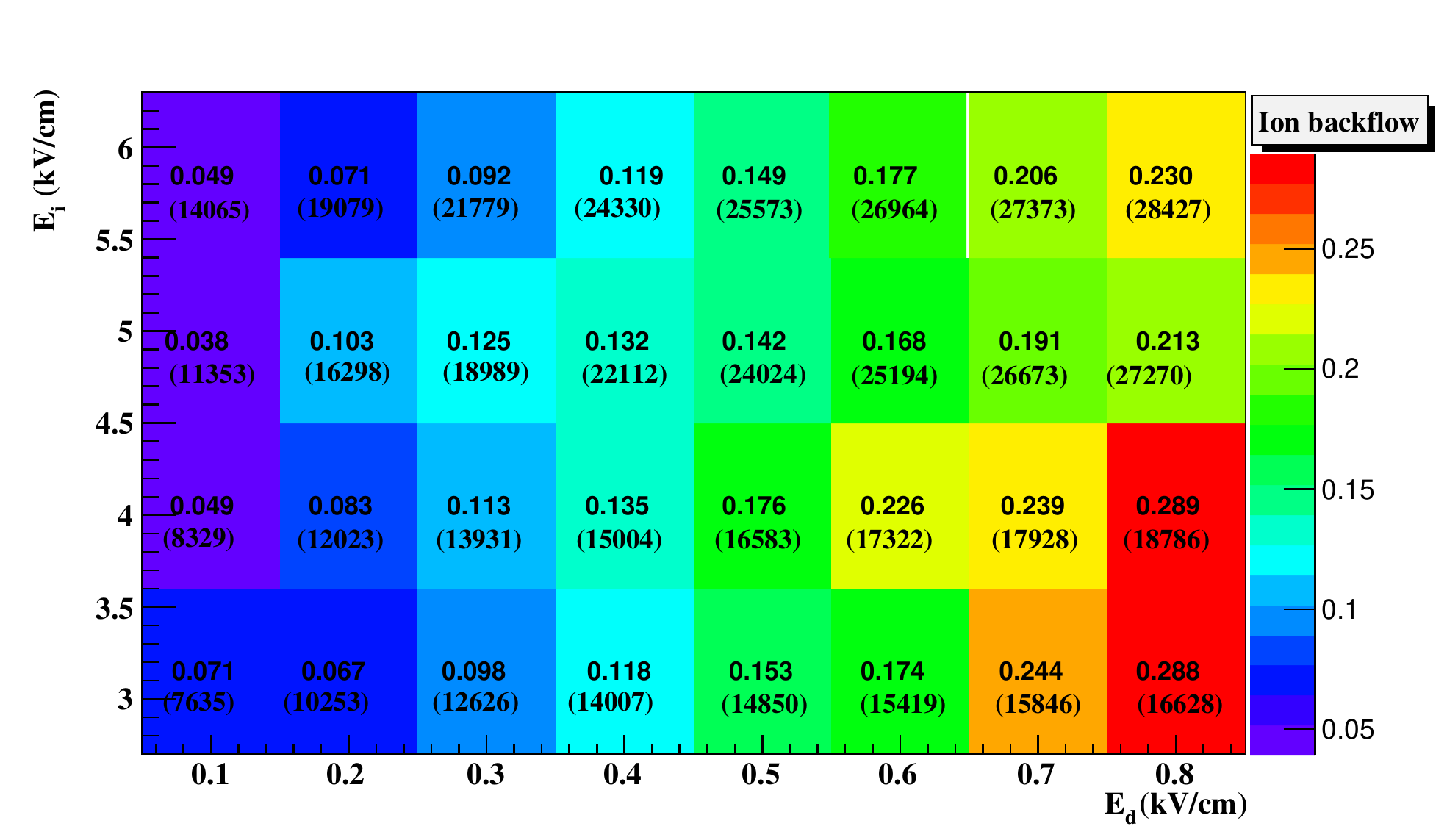}
	\caption{\label{fig:15} Ion backflow fraction of quadrupole GEM detector measured with scanning over E$_d$ and E$_i$ at constant GEM voltage $\Delta$V = 270 V for Ar/CO$_2$ 90:10. The numbers in parenthesis represent the effective gain.}
\end{figure}

First, a drift scan is done by keeping $E_i$ at 4kV/cm. The ion backflow fraction is measured with the different gas mixture and at various GEM voltages  $\Delta$V. The plots are given in Fig. \ref{fig:10} for Ar/CO$_2$ 70:30,  in Fig.  \ref{fig:11} for Ar/CO$_2$ 80:20 and in Fig. \ref{fig:12} for Ar/CO$_2$ 90:10, respectively. Here, it is observed that with increasing drift field, the ion backflow fraction increases \cite{ibf2019}, and at the highest $\Delta$V it shows a minimum value. So for the next optimisation, we have kept the GEM voltage $\Delta$V  fixed for each mixture, and a detailed study is done over drift and induction fields. The GEM voltages $\Delta$V are kept 320 V, 300 V, and 270 V for 70:30, 80:20, and 90:10 gas mixtures, respectively. The plots are given in Figs. \ref{fig:13},  \ref{fig:14},  \ref{fig:15} for gases with increasing ionisation components. 
In these figures, It is observed that for any fixed value of $E_i$, the ion backflow value increases with increasing values of $E_d$. 
Therefore, the low ion backflow fraction values are observed from low drift regions as was also recorded a similar trend from a single GEM detector. 
With increasing induction field $E_i$, which is responsible for high gain, the ion backflow fraction ratio for the quadruple GEM system is changing from high to low values. For example in Fig. \ref{fig:13}
for 70:30 Ar/CO$_2$ gas ratio with $\Delta$V=320 V, the ion backflow fraction decreases with increasing induction field $E_i$ for fixed drift field $E_d$. The minimum ion backflow fraction value is calculated and it is found to 3.5$\%$ with drift field 0.1kV in Ar/CO$_2$ in 70:30 gas mixture. 
%
%
In this case, it is observed that the more the gain, the less the ion backflow fraction.  Similarly,  in Figs. \ref{fig:14} and \ref{fig:15}, the ion backflow fraction are 3.0$\%$ and 3.8$\%$ with drift field 0.1kV in Ar/CO$_2$ mixed gas mixture 80:20 with $\Delta$V= 300 V and  90:10 with $\Delta$V =270 V, respectively. 
%
 However, for Fig. \ref{fig:13}, the ion backflow fraction changes rapidly due to a change of effective gain as a function of $E_i$. For example, the ion backflow value becomes almost half at fixed $E_d$=0.1 kV/cm from $E_i$ =3 to 5 kV/cm, due to effective gain changes from 3000 to 11000. 
 %
 %
 The higher the effective gain value  the smaller the ion backflow fraction. Similar trends are also seen in Ref.\cite{sauli2016,ibf2019, ibf2014, ibf2004}.

\section{Conclusion}

The performance in the GEM mainly depends on field configurations and gas mixtures. So a detailed measurement is done for the study of ion backflow fraction with GEM-based detectors. The effective gain and ion backflow fraction are determined for a single and quadruple GEM detector. 
Throughout the experiment, the ambient parameters like gas flow rate, temperature, pressure, and relative humidity are maintained constant. 
The dependencies of ion backflow fraction are carefully observed with E$_d$, E$_i$, $\Delta$V, and with three mixed Ar/CO$_2$  gas 70:30, 80:20 and 90:10 ratios. It is found that the ion backflow fraction is 
dependent on the configuration of the drift field and the GEM voltage $\Delta$V. The ion backflow fraction values decrease
as a function of the induction field keeping drift field as constant. With an increase in gain, the ion backflow fraction decreases proportionately. From the scanning over gas ratios, we also observed that with increasing quencher proportion, the Effective Gain decreases and ion backflow fraction value is quite high. We found a minimum ion backflow fraction of ~3.5$\%$, 3.0$\%$ and 3.8$\%$ with E$_d$ = 0.1 kV/cm for 70:30, 80:20 and 90:10 Ar/CO$_2$ gas ratios, 
respectively, for quadrupole GEM detector and a minimum ion backflow fraction of 18 $\%$ is achieved in the single GEM detector with Ar/CO$_2$ 80:20 gas mixture. 

The Effective Gain of the quadruple GEM is measured by estimating the currents from the pulse height spectrum using MCA and found to be similar to that estimated by the current measurement method.



\end{document}